\documentclass[runningheads]{llncs}
\usepackage{graphicx}
\usepackage{amssymb}
\usepackage{amsmath}

\usepackage{amsthm}
\usepackage{mathrsfs}
\usepackage{multirow}
\usepackage{xhfill}
\usepackage[hidelinks,colorlinks=true,allcolors=blue]{hyperref}
\usepackage{cleveref}
\usepackage{bm}
\usepackage[T1]{fontenc}
\usepackage{polski}
\usepackage[english]{babel}
\usepackage[inline]{enumitem}
\usepackage[title]{appendix}
\usepackage{tikz}

\newcommand{\E}{\mathtt{E}}
\newcommand{\T}{\mathcal{T}}

\newcommand{\Lan}{\mathscr{L}}
\newcommand{\M}{\mathscr{M}}
\newcommand{\D}{\mathscr{D}}
\newcommand{\I}{\mathscr{I}}
\newcommand{\Log}{\mathsf{L}}
\newcommand{\PFL}{\textsf{PFL}}
\newcommand{\NFL}{\textsf{NFL}}
\newcommand{\PQFL}{\textsf{PQFL}}
\newcommand{\NQFL}{\textsf{NQFL}}
\newcommand{\NQFLm}{\textsf{NQFL}\textsuperscript{--}}
\newcommand{\TC}{\textsf{TC}}
\newcommand{\Tab}{\mathcal{T}}
\newcommand{\B}{\mathcal{B}}

\makeatletter
\newtheorem*{rep@theorem}{\rep@title}
\newcommand{\newreptheorem}[2]{%
	\newenvironment{rep#1}[1]{%
		\def\rep@title{#2 \ref{##1}}%
		\begin{rep@theorem}}%
		{\end{rep@theorem}}}
\makeatother

\newreptheorem{proposition}{Proposition}

\newtheorem*{proposition*}{Proposition}

\newreptheorem{lemma}{Lemma}

\newtheorem*{lemma*}{Lemma}

\begin{document}
\title{Tableaux for Free Logics with Descriptions\thanks{Both authors are supported by the National Science Centre, Poland (grant number: DEC-
		2017/25/B/HS1/01268). The second author is supported by the EPSRC projects OASIS
		(EP/S032347/1), AnaLOG (EP/P025943/1), and UK FIRES
		(EP/S019111/1), the SIRIUS Centre for Scalable Data Access,
		and Samsung Research UK.}}
%
%
\author{Andrzej Indrzejczak\inst{1}\orcidID{0000-0003-4063-1651} \and
Michał Zawidzki\inst{2,1}\orcidID{0000-0002-2394-6056}}
\authorrunning{A. Indrzejczak and M. Zawidzki}
%
\institute{Department of Logic, University of Łódź, Poland \and
Department of Computer Science, University of Oxford, UK\\
\email{andrzej.indrzejczak@filhist.uni.lodz.pl}\\
\email{michal.zawidzki@cs.ox.ac.uk}}
\maketitle              
\begin{abstract}

The paper provides a tableau approach to definite descriptions. We focus on several formalizations of the so-called minimal free description theory (MFD) usually formulated axiomatically in the setting of free logic. 
We consider five analytic tableau systems corresponding to different kinds of free logic, including the logic of definedness applied in computer science and constructive mathematics for dealing with partial functions (here called \emph{negative quasi-free logic}). 
The tableau systems formalise MFD based on \PFL{} (positive free logic), \NFL{} (negative free logic), \PQFL{} and \NQFL{} (the quasi-free counterparts of the former ones). Also the logic \NQFLm{} is taken into account, which is equivalent to \NQFL{}, but whose language does not comprise the existence predicate. 
It is shown that all tableaux are sound and complete with respect to the semantics of these logics.

\end{abstract}

\setcounter{footnote}{0}
\section{Introduction}\label{sect::Introduction}

The topic of \emph{definite descriptions} (DD) is of wide interest to philosophers, linguists, and logicians. On the other hand, in proof theory and automated deduction the number of formal systems and studies of their properties is relatively modest. In particular, there are several
tableau calculi due to Bencivenga, Lambert and van Fraassen~\cite{ben:log91}, Gumb~\cite{gum:des00},~Bostock \cite{bos:int00}, Fitting and Mendelsohn~\cite{fitt:fir98}, but all of them introduce DD by means of rather complex rules, and so, are not really in the spirit of tableau methodology. Quite a lot of
natural deduction systems for DD have been provided, but only a few of them (namely Tennant's~\cite{ten:des01,ten:des02} and K\"urbis' \cite{kur:bsl19,kur:bsl20} works) deal with DD by means of rules which allow for finer proof analysis and provide normalization proofs. Cut-free sequent calculi for several theories of DD were provided by Indrzejczak~\cite{ind:aml18,ind:llp18,ind:aml20,ind:llp20} and recently also by Orlandelli \cite{orlandelli}.  

The number of theories of DD that have been proposed since Frege's and Russell's first accounts (see, e.g., a discussion in~\cite{pel:lin00}) is enormous, however
what we are concerned with in this paper is an adequate tableau characterization of DD, so due to space restrictions we omit a detailed presentation of different theories of DD and their philosophical or linguistic motivations. In particular, we confine ourselves to only one approach to DD, strongly connected with \emph{free logic} and commonly called a \emph{minimal free description theory} (\textsf{MFD})\footnote{The reader may find a more fine-grained presentation of \textsf{MFD} and its extensions in Lambert's \cite{lam:des01}, Bencivenga's \cite{benc:free86} or Lehmann's \cite{leh:fre00} works.}. It is based on the so-called Lambert's axiom (\textsf{L}):
\begin{align}
\forall x(\imath x \varphi(x)=x \leftrightarrow \forall y(\varphi(y) \leftrightarrow y=x)).\tag{\textsf{L}}\label{cond::Lambert} 
\end{align}
In fact, this axiom added to different kinds of free logics leads to significantly different theories of DD. We provide tableau calculi for four kinds of different free logics, called here \PFL, \NFL, \PQFL, and \NQFL{} (where \textsf{N} stands for \emph{negative}, \textsf{P} for \emph{positive}, \textsf{Q} for \emph{quasi}). 
In negative free logics, in contrast to positive ones, atomic formulas with non-denoting terms are always evaluated as false or, equivalently, all predicates are strict, that is, defined only over denoting terms.  
Both \PFL{} and \NFL{} characterize absolutely free logics in the sense that variables may also fail to denote. On the other hand,
\NQFL{} and \PQFL{} are systems for quasi-free logics in the sense that only descriptions can fail to denote; variables are always denoting.

Recently, cut-free sequent calculi for several free logics, yet without DD, have been presented by Pavlović and Gratzl~\cite{pavl:21} and  by
 Indrzejczak \cite{ind:des20}. In particular, in the latter work it has been shown that if we restrict instantiation in quantifier rules only to variables, we do not lose completeness, provided that some special rules are added. It makes it possible to characterize \NQFL{} and \PQFL{} by means of classical quantifier rules, which justifies our use of the term `quasi free' (introduced therein).Yet even more importantly, such a restriction on quantifier rules allows us 
 to extend this approach to \textsf{MFD} and preserve cut-freeness (see \cite{ind:llp20}). Since the above-referenced paper provides a purely proof-theoretic approach, completing the work with the semantic side and suitably defined adequate and analytic tableau systems seems to be a natural next research step. The aim of the present study is to make this step and fill the indicated gap.

We limit our considerations to the logics mentioned above as the most prominent representatives of the family of free logics.
\PFL{} is by all means the most popular version of free logic (see, e.g., \cite{benc:free86}, \cite{lam:des02}, or \cite{leh:fre00}), applied mainly in philosophical studies and as the basis of formalization of modal first-order logics (see, e.g., Garson \cite{gars:foml06}). The original Lambert's version of \textsf{MFD} was proposed on the basis of \PFL.
The basic negative free logic \NFL, known also as the \emph{logic of existence} (\cite{sco:fre00}), was more popular in computer science and foundational studies~\cite{ten:des01,ten:des02}. 

Negative quasi-free logic \NQFL{} is known as the \emph{definedness logic} (or the \emph{logic of partial terms}) by Beeson~\cite{bee:fre00} and Feferman~\cite{fef:fre00}. It has also been extensively studied and applied in computer science. Although it was originally developed in the context of constructive mathematics to deal with partial untyped combinatory and lambda calculi, Feferman rightly noticed that it works without changes in the classical setting (in fact, he was concerned only with classical semantics in \cite{fef:fre00}. \PQFL{} is a positive variant of \NQFL{}, that is, not requiring that all predicates are strict. It is interesting that its intuitionistic restricted version (no identity and DD) was studied proof-theoretically by Baaz and Iemhoff~\cite{baa:fre00} and recently by Maffezioli and Orlandelli~\cite{maf:fre00}. 

\NQFLm{} is a variant of \NQFL{} but formulated in the language without the existence predicate. Although the latter can be defined in all the considered logics, it is handy to keep it as primitive. However, in~\cite{ind:des20} it was shown that in quantifier rules for all free logics with identity, instantiation terms may be restricted to variables. That opens a possibility of discarding the existence predicate and simplifying the rules, at least for \NQFL. Thus, this logic is presented here in two variants: as \NQFL{} with the existence predicate (which allows to compare it with the remaining logics more easily), and then as \NQFLm{} in an existence-free version with simpler rules. In fact \NQFLm{} with the rules for descriptions on classical foundations appears to be equivalent also to the formalization of Russellian theory of descriptions provided by Kalish, Montague and Mar~\cite{kali:log64}; (see Indrzejczak~\cite{ind:des21} for a detailed explanation).

Lambert's axiom (\ref{cond::Lambert}) was used as a basic way of formalizing DD in all the abovementioned logics, except for \PQFL. However, on the ground of \NFL, (and \NQFL) it yields quite a strong theory of DD of essentially Russellian character. This follows from the fact that in \NFL{} (\NQFL) (\ref{cond::Lambert}) is equivalent to the following formula:
\begin{align}
\psi(\imath x\varphi(x)) \ \leftrightarrow \ \exists y(\forall x(\varphi(x) \leftrightarrow x = y)\wedge\psi(y)),\text{ where $\psi$ is atomic}.\tag{$\textsf{R}$}\label{cond::Russell}
\end{align}
(\ref{cond::Russell}) expresses the Russellian approach to characterizing DD and it was often attacked as being too strong. The left-to-right implication encodes that if we state something about a DD, it implies that this description denotes. According to Strawson's well-known criticism, if a DD is used as an argument of a predicate, its existence and uniqueness is presupposed rather than implied.    
Lambert's axiom is in general weaker than (\ref{cond::Russell}) and in \PFL{} (\PQFL{}) implies only the right-to-left implication of (\ref{cond::Russell}) which is commonly acceptable. The equivalence of (\ref{cond::Lambert}) and (\ref{cond::Russell}) in \NFL{} is a consequence of the fact that in \NFL{} all predicates are strict, so the statement of an atomic formula implies that all terms occurring in it are denoting (see \cite{ind:llp20}).

Due to space limitations, we confine ourselves to logics which are founded on the classical core. Interestingly, cut-free sequent calculi in~\cite{ind:llp20}, after restricting sequents to at most one formula in the succedent and small refinements of some rules for DD, may also characterize their intuitionistic versions. In the case of tableaux adequate with respect to a given semantics, however, such small refinements do not suffice to obtain intuitionistic versions. Hence, we postpone completing this task, as well as the characterization of \textsf{MFD} on the basis of neutral free logics, to future work. In the latter case even the standard sequent calculus is not sufficient for a satisfactory proof-theoretic characterization. 

In what follows, after a brief characterization of the syntax and semantics in Section~\ref{sect::Preliminaries}, in Section~\ref{sect::TableauCalculi} we provide five tableau calculi for the logics \PFL, \PQFL, \NFL, \NQFL, and \NQFLm. Adequacy of all systems is established in Section~\ref{sect::SoundnessCompleteness}. In Section~\ref{sect::Related} we briefly compare our tableau calculi with alternative approaches, in particular with sequent calculi by Indrzejczak~\cite{ind:llp20}. Finally we discuss some possible advantages of using DD instead of functional terms and present further lines of research.

\section{Preliminaries}\label{sect::Preliminaries}

\subsection{Syntax}
For the logics \PFL, \NFL, \PQFL, \NQFL{} we consider sentences, that is, formulas with no free variables, built in the standard  first-order language $\Lan$ with identity and the unary existence predicate $\E$ treated as logical constants and with no function symbols as primitives. The vocabulary of $\Lan$ consists of:
\begin{itemize}
	\item a countably infinite set of bound individual variables $\mathsf{VAR}=\{ x, y, z\ldots \}$,
	\item a countably infinite set of parametric (free) individual variables $\mathsf{PAR}=\{ a, b, c,\ldots\}$,
	\item a countably infinite set of $n$-ary predicate symbols $\mathsf{PRED}^n = \{P^n, Q^n, R^n, \ldots\}$, for any non-negative integer $n$;
	\item a set of propositional connectives: $\neg$, $\land$,
	\item the universal quantifier $\forall$,
	\item the definite description operator $\imath$,
	\item the identity relation $=$,
	\item the existence predicate $\E$,
	\item left and right parentheses: (, ).
\end{itemize}
In the case of \NQFLm we discard the existence predicate $\E$ from the language and refer to such a restricted language as $\Lan^-$.

A set of terms $\mathsf{TERM}$ and a set of formulas $\mathsf{FOR}$ (in the language of deduction) are defined simultaneously by the following context-free grammars:
\begin{gather*}
\mathsf{TERM} \ni t ::= x \mid a \mid \imath x \varphi,\\
\mathsf{FOR} \ni \varphi ::= P(t_1 ,\ldots, t_n) \mid t_1 = t_2 \mid \E t \mid \neg \varphi \mid \varphi \land \varphi \mid \forall x \varphi,
\end{gather*}
where $x \in \mathsf{VAR}$, $a \in \mathsf{PAR}$, $P \in \mathsf{PRED}^n$, $t, t_1, \ldots, t_n \in \mathsf{TERM}$, and $\varphi \in \mathsf{FOR}$. The existential quantifier and other boolean connectives are introduced as standard abbreviations.
Note that the absence of function symbols as primitives in $\Lan$ and $\Lan^-$ is due to the fact that they can be simulated by using the operator $\imath$ in the sense that every term of the form $f^n(t_1, \ldots, t_n)$ can be represented as $\imath xF^{n+1}(t_1, \ldots, t_n, x)$. On the other hand, not every (proper) description can be expressed using functional terms. For example, descriptions like `the winner of the ultimate fight', `the bear we have seen recently' can only be represented by constants.

\subsection{Semantics}\label{subsect::Semantics}
	
	By a \emph{model} we mean a structure $\M=\langle \D, \D_\E, \I \rangle$, where $\D_\E$ is a (possibly empty) subset of $\D$ and for each $n$-argument predicate $P^n$, $\I(P^n)\subseteq \D^n$. An \emph{assignment} $v$ is defined as $v:VAR\cup PAR\longrightarrow \D$ for \PFL, \NFL, and as
	$v:VAR\cup PAR\longrightarrow \D_\E$ for \PQFL, \NQFL, and \NQFLm. Thus, in proper free logics variables may fail to denote, which is not possible in quasi-free logics.
	An \emph{$x$-variant} $v'$ of $v$ agrees with $v$ on all arguments, save, possibly, $x$. We will write $v^x_o$ to denote the $x$-variant of $v$ with $v^x_o(x) = o$. The notion of \emph{interpretation} $\I_v (t)$ of a term $t$ under an
	assignment $v$ is defined simultaneously with the notion of \emph{satisfaction} of a formula $\varphi$ under $v$, in symbols $\M, v \models \varphi$:\medskip
	
	\begin{tabular}{rcl}
		
		$\I_v (x) = v(x)$, & &\\[.5ex]
		$\I_v (a) = v(a)$, & &\\[.5ex]
		$\I_v(\imath x\varphi) = o\in \D_\E$  & \parbox{1.5cm}{\centering iff}  &  \begin{minipage}[t]{6.5cm}$\M, v^x_{o} \models \varphi$, and for any $x$-variant $v'$ of $v$, if $\M, v' \models \varphi$, then $v'(x)=o$,\end{minipage}\\[.5ex]
		$\M, v \models P^n(t_1, ..., t_n) $ & iff & \begin{minipage}[t]{6.5cm}$\langle \I_v(t_1), \ldots, \I_v(t_n) \rangle \in \I (P^n)$ \\
		(and $\I_v(t_i)\in \D_\E, i\leq n$, for \NFL{}, \NQFL, and \NQFLm),\vspace{.4\baselineskip}\end{minipage}\\
		$\M, v \models t_1 = t_2 $ & iff & \begin{minipage}[t]{6.5cm}$\I_v(t_1) = \I_v(t_2)$\\
		(and $\I_v(t_1), \I_v(t_2)\!\in\! \D_\E$, for \NFL{}, \NQFL, and \NQFLm),\vspace{.4\baselineskip}\end{minipage} \\[2ex]
		$\M, v \models \E t $ & iff & $\I_v(t)\in \D_\E$,\\[.5ex]
		$\M, v \models \neg \varphi$ & iff & $\M, v \not\models \varphi$, \\[.5ex]	
		$\M, v \models \varphi \land \psi $ & iff & $\M, v\models \varphi$ and $\M, v \models \psi$,\\[.5ex]
		$\M, v \models \forall x\varphi $ & iff & $\M, v^x_o \models \varphi$, for all $o\in \D_\E$,
	\end{tabular}\\[\baselineskip]
where $x \in \mathsf{VAR}$, $a \in \mathsf{PAR}$, $P^n \in \mathsf{PRED}^n$, and $t, t_1, \ldots, t_n \in \mathsf{TERM}$.

	A formula $\varphi$ is called \emph{satisfiable} if there exist a model $\M$ and a valuation $v$ such that $\M, v \models \varphi$. A formula is \emph{valid} if, for all models $\M$ and valuations $v$, $\M, v \models \varphi$. In the remainder of the paper, instead of writing $\M , v \models \varphi_1, \ldots, \M, v \models \varphi_n$, we will write $\M, v \models \varphi_1, \ldots, \varphi_n$.

\section{Tableau Calculi}\label{sect::TableauCalculi}

In this section, we present tableau calculi for the considered logics for definite descriptions. For each logic $\Log \in \{\PFL, \NFL, \PQFL, \NQFL, \NQFLm\}$ we denote the tableau calculus for $\Log$ by $\TC_\Log$.

A \emph{tableau} $\Tab$ generated by a calculus $\TC_\Log$, for $\Log \in \{\PFL, \NFL, \PQFL, \NQFL,\linebreak \NQFLm\}$, is a \emph{derivation tree} whose nodes are assigned formulas in a respective (deduction) language. A \emph{branch of $\Tab$} is a simple path from the root to a leaf of $\Tab$. For brevity, we identify each branch $\B$ with the set of formulas assigned to nodes constituting $\B$.

Our tableau calculi are composed of rules whose general form is as follows: $\frac{\Phi}{\Psi_1 | \ldots | \Psi_n}$, where $\Phi$ is the set of \emph{premises} and each $\Psi_i$, for $i\in\{1,\ldots,n\}$, is a set of \emph{conclusions}. If a rule has more than one set of conclusions, it is called a \emph{branching} rule. Otherwise it is \emph{non-branching}. Thus, if a rule $\frac{\Phi}{\Psi_1 | \ldots | \Psi_n}$ is applied to $\Phi$ occurring on $\B$, $\B$ splits into $n$ branches: $\B \cup \{\Psi_1\}, \ldots, \B \cup \{\Psi_n\}$. A rule $(\sf R)$ with $\Phi$ as the set of its premises is \emph{applicable} to $\Phi$ occurring on a branch $\B$ if it has not yet been applied to $\Phi$ on $\B$. A set $\Phi$ is called \emph{$(\sf R)$-expanded} if $(\sf R)$ has already been applied to $\Phi$.
A term $t$ is called \emph{fresh} on a branch $\B$ if it has not yet occurred on $\B$. We call a branch $\B$ \emph{closed} if the inconsistency symbol $\bot$ occurs on $\B$. If $\B$ is not closed, it is \emph{open}. A branch is \emph{fully expanded} if it is closed or no rules are applicable to (sets of) formulas occurring on $\B$. A tableau $\Tab$ is called closed if all of its branches are closed. Otherwise $\Tab$ is called open. Finally, $\Tab$ is fully expanded if all its branches are fully expanded.
A \emph{tableau proof} of a formula $\varphi$ is a closed tableau with $\neg\varphi$ at its root. A formula $\varphi$ is tableau-valid (with respect to the calculus $\TC_\Log$) if all fully expanded tableaux generated by $\TC_\Log$ with $\neg\varphi$ at the root are tableau proofs of $\varphi$.
A tableau calculus $\TC_\Log$ is \emph{sound} if, for each formula $\varphi$, whenever $\varphi$ is tableau-valid wrt $\TC_\Log$, then it is valid. $\TC_\Log$ is \emph{complete} if, for each formula $\varphi$, whenever $\varphi$ is valid, then it is tableau-valid wrt $\TC_\Log$. 

When presenting the rules, we adopt the following notational convention:
\begin{itemize}
	\item metavariables $\varphi$, $\psi$ stand for arbitrary formulas in $\Lan$ (or $\Lan^-$ if \NQFLm is considered),
	\item metavariables $t, t_1, \ldots, t_n$ represent arbitrary terms present on a branch,
	\item metavariables $a$, $a_1,\ldots, a_n$ denote fresh parameters,
	\item metavariables $b$, $b_1$, $b_2$ stand for an arbitrary parameters present on a branch,
	\item an expression $\varphi[x/t]$ represents the result of a correct substitution of all free occurrences of $x$ within $\varphi$ with a term $t$,
	\item $t_1 \neq t_2$ is an abbreviation for $\neg(t_1 = t_2)$,
	\item `DD' is an abbreviation for `definite description'.
\end{itemize}

The rules for tableau calculi $\TC_\PFL$, $\TC_\NFL$, $\TC_\PQFL$, $\TC_\NQFL$, and $\TC_\NQFLm$ are presented in Figures~\ref{fig::Rules} and \ref{fig::TableauCalculi}. Intuitively, if a rule's name contains `$E$' and the name of an operator, it is an elimination rule which removes the operator from the processed formula. On the other hand, if a rule's name contains `$I$' and the name of an operator, it is an \emph{introduction} rule which adds to the branch an expression featuring this operator. Moreover, we have three \emph{closure} rules which close the branch as inconsistent, and two special \emph{analytic cut} rules which make it possible to compare denotations of variables and definite descriptions.

\begin{figure}[t!]
	\centering
	\textbf{Rules}\smallskip
	
	\parbox{0.75cm}{CPL} \xrfill{0.5pt}\medskip
	
	$(\neg\neg E)$\ $\dfrac{\neg\neg \varphi}{\varphi}$\qquad\
	$(\land E)$\ $\dfrac{\varphi \land \psi}{\varphi, \psi}$\qquad\ $({\neg\land} E)$\ $\dfrac{\neg(\varphi \land \psi)}{\neg\varphi \mid \neg\psi}$\medskip
	
	\parbox{0.75cm}{$\bot$} \xrfill{0.5pt}\medskip
	
	$(\bot_1)$\ $\dfrac{\varphi, \neg\varphi}{\bot}$\qquad	$(\bot_2)$\ $\dfrac{t \neq t}{\bot}$\qquad $(\bot_3)$\ $\dfrac{b \neq b}{\bot}$\medskip
	
	\parbox{0.75cm}{$\forall$} \xrfill{0.5pt}\medskip
	
	$(\forall E_1)$\ $\dfrac{\forall x\varphi}{\varphi[x/b]}$\qquad
	$(\neg\forall E_1)$\ $\dfrac{\neg\forall x\varphi}{\neg\varphi[x/a]}$\qquad
	$(\forall E_2)$\ $\dfrac{\forall x\varphi, \E b}{\varphi[x/b]}$\qquad
	$(\neg\forall E_2)$\ $\dfrac{\neg\forall x\varphi}{\E a, \neg\varphi[x/a]}$\medskip
	
	\parbox{0.75cm}{$=$} \xrfill{0.5pt}\medskip
	
	$(=E)$\ $\dfrac{t_1 \approx t_2, \varphi[x/t_1]}{\varphi[x/t_2]}$,\ $t_1 \approx t_2$ stands for $t_1 = t_2$ or $t_2 = t_1$  \\[2ex]
	$(= I_1)$\ $\dfrac{P(t_1, \ldots, t_n)}{a_i= t_i}$,\ $1\!\leq\! i\!\leq\! n$ and $t_i$ is a DD\quad
	$(= I_2)$\ $\dfrac{t_1 = t_2}{a_i = t_i}$,\ $1\!\leq\! i\!\leq\! 2$ and $t_i$ is a DD\\[2ex]
	$(cut_1)$\ $\dfrac{}{b=t\mid b\neq t}$,\ $t$ is a DD\qquad $(cut_2)$\ $\dfrac{\E b}{b = t \mid b \neq t}$,\ $t$ is a DD
    \smallskip
	
	\parbox{0.75cm}{$\E$} \xrfill{0.5pt}\medskip
	
	$(\E E_1)$\ $\dfrac{\E t}{a = t}$,\ $t$ is a DD\qquad $(\E E_2)$\ $\dfrac{\E t}{t = t}$\\[2ex]
	$(\E I_1)$\ $\dfrac{P(t_1, ..., t_n)}{\E t_i}$,\ $1\!\leq\! i\!\leq\! n$ (and $t_i$ is a DD for \NFL)\\[2ex]
	$(\E I_2)$\ $\dfrac{t_1 = t_2}{\E t_i}$,\ $1\!\leq\! i\!\leq\! 2$ (and $t_i$ is a DD for \NFL)\\[2ex]
	$(\E I_3)$\ $\dfrac{}{\E b}$\qquad
	$(\E I_4)$\ $\dfrac{}{\parbox{1cm}{\centering $\E a$}}$,\ if there are no parameters on the branch\medskip
	
	\parbox{0.75cm}{$\imath$} \xrfill{0.5pt}\medskip
	

	$(\imath E_1)$\ $\dfrac{b_1=\imath x\varphi}{\varphi[x/b_1], \neg\varphi[x/b_2] \mid b_1 = b_2, \varphi[x/b_1]}$\qquad
	$(\neg\imath E_1)$\ $\dfrac{b\neq \imath x\varphi}{\neg\varphi[x/b] \mid a\neq b, \varphi[x/a]}$\\[2ex]
	
	$(\imath E_2)$\ $\dfrac{b_1=\imath x\varphi, \E b_1,\E b_2}{\varphi[x/b_1], \neg\varphi[x/b_2] \mid b_1 = b_2, \varphi[x/b_1]}$\qquad
	$(\neg\imath E_2)$\ $\dfrac{b\neq \imath x\varphi, \E b}{ \neg\varphi[x/b] \mid a\neq b, \varphi[x/a], \E a}$
	

	\caption{Tableau rules for $\TC_\PFL$, $\TC_\NFL$, $\TC_\PQFL$, $\TC_\NQFL$, and $\TC_\NQFLm$}
	\label{fig::Rules}
\end{figure}

\begin{figure}[t!]
	\centering
	\bgroup
	\def\arraystretch{1.5}
	\setlength\tabcolsep{5pt}
	\begin{tabular}{c|c|c|c|c|c|}
		\cline{2-6}
		&\PFL & \PQFL & \NFL & \NQFL & \NQFLm\\
		\cline{2-6}
		&\multicolumn{5}{|c|}{$(\neg\neg E)$, $(\land E)$, $({\neg}{\land} E)$, $(\bot_1)$, $(=E)$}\\
		\cline{2-6}
		&$(\bot_2)$ & $(\bot_2)$ & $(\E E_2)$ & $(\bot_3)$ & $(\bot_3)$\\
		&$(\forall E_2)$ & $(\forall E_1)$ & $(\forall E_2)$ & $(\forall E_1)$ & $(\forall E_1)$\\
		&$(\neg\forall E_2)$ & $(\neg\forall E_1)$ & $(\neg\forall E_2)$ & $(\neg\forall E_1)$ & $(\neg\forall E_1)$\\
		&$(cut_2)$&$(cut_2)$&$(cut_2)$&$(cut_2)$&$(cut_1)$\\
		&$(\E E_1)$ & $(\E E_1)$ & $(\E E_1)$ & $(\E E_1)$ &\\
		& && $(\E I_1)$ & $(\E I_1)$ & $(= I_1)$\\
		&&& $(\E I_2)$ & $(\E I_2)$ & $(= I_2)$\\
		&&$(\E I_3)$&&$(\E I_3)$&\\
		&$(\imath E_2)$ & $(\imath E_1)$ & $(\imath E_2)$ & $(\imath E_1)$ & $(\imath E_1)$\\
		&$(\neg\imath E_2)$ & $(\neg\imath E_1)$ & $(\neg\imath E_2)$ & $(\neg\imath E_1)$ & $(\neg\imath E_1)$\\
		\hline
		\multicolumn{1}{|c|}{\begin{minipage}{3cm}\centering non-empty domain assumption\end{minipage}} & $(\E I_4)$ && $(\E I_4)$ &&\\
		\hline
	\end{tabular}
	\egroup	
	\caption{Tableau calculi $\TC_\PFL$, $\TC_\NFL$, $\TC_\PQFL$, $\TC_\NQFL$, and $\TC_\NQFLm$}
	\label{fig::TableauCalculi}
\end{figure}

A few words of comment on the rules displayed in Figure~\ref{fig::Rules} are in order. The propositional core of the calculi is known from tableaux for classical propositional logic. The rule $(\bot_1)$ closes a branch when a propositional inconsistency occurs thereon, whereas the remaining two closure rules, $(\bot_2)$ and $(\bot_3)$ rest on reflexivity of identity (possibly in a restricted form). The rules $(\forall E_1)$ and $(\neg\forall E_1)$ are standard rules for quantifier elimination in first-order logic. The remaining two rules for $\forall$, namely $(\forall E_2)$ and $(\neg\forall E_2)$, reflect the semantic condition saying that a term replacing a variable after quantifier elimination must denote an existing object. While in quasi-free logics it is ensured by the definition of valuation, in the remaining (absolutely free) logics it needs to be secured by a separate existence formula. Note that all quantifier elimination rules admit only parameters as instances of bound variables. The $(= E)$-rule scheme ensures the substitutability of identical terms within arbitrary formulas, often called Leibniz' principle. One of its side effects is a guarantee that $=$ is symmetric in all calculi. $(= I_1)$ and $(= I_2)$, occurring only in $\TC_\NQFLm$, which lacks the existence predicate $\E$, make sure that each definite description occurring in a true atomic formula has a unique and existing denotation, by equating it with a fresh variable (which is always denoting in $\NQFLm$). $(cut_1)$ and $(cut_2)$ are a restricted form of analytic cut which, for each definite description and denoting variable checks whether their denotations are identical or distinct. $(\E E_1)$ works similarly to $(= I_1)$ and $(= I_2)$ with the caveat that it equates with a fresh variable a definite description that is known to be denoting. $(\E E_2)$, which is present only in $\TC_\NFL$, enforces reflexivity of identity among denoting terms. Intuitively, it allows us to prove that, for each non-denoting term $t$, a formula $t\neq t$ holds in \NFL. The rules $(\E I_1)$ and $(\E I_2)$ reflect the semantic condition stating that each term which is an argument of a true atomic \NQFL-formula, or each definite description occurring in such an \NFL-formula, is denoting. $(\E I_3)$, on the other hand, refers to the definition of valuation in \PQFL{} and \NQFL, where variables are always mapped to existing objects. The rule $(\E I_4)$ introduces a fresh variable which is assumed to denote, provided that there are no parameters on the branch. Consequently, it guarantees that the non-empty domain assumption is satisfied, should we make it.
The first pair of $\imath$-rules, $(\imath E_1)$ and $(\neg \imath E_1)$, eliminate an occurrence of a definite description provided that it appears as an argument of an identity. In $(\imath E_1)$ a formula defining the definite description must hold of $b_1$, hence this formula is present in both conclusions. A definite description is subsequently compared to each parameter $b_2$ occurring on a branch. If we assume that they are equal, it is also equal to $b_1$ (the right conclusion), otherwise $\varphi$ does not hold of $b_2$, so we obtain its negation. In $(\neg\imath E_1)$ we assume that a denoting parameter $b$ and a definite description have distinct denotations. It is either because the formula defining the definite description does not hold of $b$ (the left conclusion) or because some other object satisfies this formula. To state the latter a fresh parameter $a$ is introduced which satisfies $\varphi$, yet it is not equal to $b$. The second pair of $\imath$-rules, $(\imath E_2)$ and $(\neg\imath E_2)$, being a part of the calculi for proper free logics, work similarly, with the caveat that we need to additionally ensure, using the existence predicate $\E$, that respective variables occurring in the premises of the rules are denoting. In \PFL{} and \NFL{} variables are not automatically guaranteed to denote, so such an additional condition is necessary for bringing the rules in line with the semantic condition for proper definite descriptions.

Since the rules in all calculi are closed under subformulas modulo substitution, adding single negations and adding equality to two terms already present on the branch one of which being a definite description and another one being a parameter, one can think of the calculi as \emph{analytic} in an extended sense of the term.

\section{Soundness and Completeness}\label{sect::SoundnessCompleteness}
In order to prove soundness and completeness of the calculi $\TC_\PFL$, $\TC_\PQFL$, $\TC_\NFL$, $\TC_\NQFL$, and $\TC_\NQFLm$ we need two well-known lemmas which we recall without proofs (see, e.g.,~\cite[Sect. III.4 and III.8]{EbFlTh:96}).

\begin{lemma}[Coincidence Lemma]\label{lem::Coincidence}
	Let $\varphi \in \mathsf{FOR}$, let $\M = \langle \D,\D_\E,\I\rangle$ be a model, and let $v_1,v_2$ be assignments. If $v_1(x)=v_2(x)$ for each free variable $x$ occurring in $\varphi$, then $\M, v_1 \models \varphi$ iff 
		$\M, v_2 \models \varphi$.
\end{lemma}

\begin{lemma}[Substitution Lemma]\label{lem::Substitution}
	Let $\varphi \in \mathsf{FOR}$, $t,t'\in\mathsf{TERM}$, and let $\M = \langle \D,\D_\E,\I\rangle$ be a model. Then 
	$\M, v \models \varphi[x/t]$ iff $\M, v^x_{\I_v(t)}\models \varphi$.
\end{lemma}

\subsection{Soundness}\label{subsect::Soundndess}

Let $(\mathsf{R})$ $\frac{\Phi}{\Psi_1\mid\ldots\mid\Psi_n}$ be a rule from a calculus $\TC_\Log$. We say that $(\mathsf{R})$ is \emph{sound} if whenever $\Phi$ is $\Log$-satisfiable, then $\Phi\cup\Psi_i$ is $\Log$-satisfiable, for some $i \in \{1,\ldots,n\}$.

\begin{lemma}\label{lem::SoundRules}
	For each $\Log \in \{\PFL, \PQFL, \NFL, \NQFL, \NQFLm\}$ all rules of $\TC_\Log$ are sound.
\end{lemma}

\begin{proof}
	We confine ourselves to showing soundness of the rules for definite descriptions. The proof of the remaining cases can be found in the \hyperref[sect::appendix]{Appendix}.\smallskip
	
	\noindent To prove soundness of $(\imath E_1)$ assume that $b_1 = \imath x \varphi$ is $\Log$-satisfiable, for $\Log \in \{\PQFL, \NQFL, \NQFLm\}$, that is, there exists a model $\M = \langle \D, \D_\E, \I \rangle$ and an assignment $v$ such that $\M, v \models b_1 = \imath x \varphi$. Let $v(b_1)= o\in \D_\E$, then $\I_v(\imath x \varphi) = v(b_1) = o$ and by the satisfaction condition $\M, v^x_{o} \models \varphi$, and for any $x$-variant $v'$ of $v$, if $\M, v' \models \varphi$, then $v'(x)=o$. The first conjunct guarantees, by \href{lem::Substitution}{Substitution Lemma}, that $\M, v \models \varphi[x/b_1]$, which holds for both conclusions. The second conjunct yields, for any $b_2\in \D_\E$, that either $\M , v \not \models \varphi[x/b_2]$ or $\M , v \models b_1 = b_2$. The former case yields the left conclusion, whereas the latter case yields the right one.
	To show that $(\neg \imath E_1)$ is sound assume that $b \neq \imath x  \varphi$ is $\Log$-satisfiable for $\Log \in \{\PQFL, \NQFL, \NQFLm\}$. Then, there exists a model $\M = \langle \D, \D_\E, \I \rangle$ and an assignment $v$ such that $\M , v \models b \neq \imath x \varphi$. It means that $\I_v(\imath x \varphi) \neq v(b)=o \in \D_\E$. By the satisfaction condition $\M, v^x_{o} \not \models \varphi$, or for some $x$-variant $v'$ of $v$, $\M, v' \models \varphi$ but $o'=v'(x)\neq v(x)=o$. In the first case, by \href{lem::Substitution}{Substitution Lemma}, $\M, v \not \models \varphi[x/b]$, so the left conclusion is satisfied. If the second holds, then by \href{lem::Coincidence}{Coincidence Lemma} and \href{lem::Substitution}{Substitution Lemma} we have that $\M, v \models \varphi[x/a]$ but  $\M , v \models b \neq a$ for some fresh $a$.\smallskip
	
	\noindent Proofs for $(\imath E_2)$ and $(\neg \imath E_2)$, respectively, are conducted analogically with the following caveat. In \PFL{} and \NFL{} variables are not automatically guaranteed to denote, so the existence of a referrent object needs to be ensured externally. This is done by placing a variable in the scope of the existence predicate~$\E$.
\end{proof}

Now we are ready to prove the following theorem.

\begin{theorem}[Soundness]\label{thm::Soundness}
	The tableau calculi $\TC_\PFL$, $\TC_\PQFL$, $\TC_\NFL$,\linebreak $\TC_\NQFL$, and $\TC_\NQFLm$ are sound.
\end{theorem}

\begin{proof} To show that for each $\Log$-formula $\varphi$, where $\Log \in \{\PFL, \PQFL, \NFL, \NQFL,\linebreak \NQFLm\}$, if $\varphi$ is tableau-valid, then it is valid. Let $\Tab$ be a proof of $\varphi$, that is, a closed tableau with $\neg \varphi$ at the root. Each branch of $\Tab$ has $\bot$ at the leaf, which is clearly $\Log$-unsatisfiable. By \Cref{lem::SoundRules} we know that all the rules of $\TC_\Log$ are $\Log$-satisfiability preserving, and so, going from the bottom to the top of $\Tab$, at each node we have an $\Log$-unsatisfiable set of formulas. Thus, (a singleton set consisting of) $\neg\varphi$ is $\Log$-unsatisfiable. By the well known duality between satisfiability and validity we obtain that $\varphi$ is $\Log$-valid.
\end{proof}

\subsection{Completeness}

In this section, we prove that, for each $\Log \in \{ \PFL, \PQFL, \NFL, \NQFL, \NQFLm\}$, $\TC_\Log$ is complete. To that end we show that every open and fully expanded branch $\B$ of a $\TC_\Log$-tableau $\Tab$ satisfies some syntactic conditions. Then we show how to construct an $\Log$-structure $\M_\B^\Log$ and a function $v_\B^\Log$ out of such an open and fully expanded branch, and show that $v_\B^\Log$ is an $\Log$-valuation, and $\M_\B^\Log$ is an $\Log$-model satisfying, for each $\Log$-formula $\varphi$ occurring on $\B$, $\M_\B^\Log , v_\B^\Log \models \varphi$.

We assume that for each $\Log \in \{\PFL, \PQFL, \NFL, \NQFL, \NQFLm\}$, the calculus $\TC_\Log$ can be accompanied by a suitable \emph{fair} procedure in the sense that whenever a rule can be applied, it will eventually be applied. For example, an algorithm from~\cite{fitt:fir98}, with added steps for additional rules, can be applied to $\TC_\Log$. Thus, a fully expanded, possibly infinite, branch $\B$ is \emph{closed under rule application}.

Let $\B$ be an open and fully expanded branch of a $\TC_\Log$-tableau $\Tab$, where $\Log \in \{ \PFL, \PQFL, \NFL, \NQFL, \NQFLm\}$.  Let $\mathsf{TERM}(\B)$, $\mathsf{VAR}(\B)$, and $\mathsf{PAR}(\B)$ be the sets of, respectively, all terms occurring on $\B$ (that is, parameters and definite descriptions), all bound variables occurring on $\B$, and all parameters occurring on $\B$. We define a binary relation $\sim$ on $\mathsf{TERM}(\B)$ in the following way:
$$\forall t_1 , t_2 \in \mathsf{TERM}(\B)\quad \big[t_1\sim t_2\quad\text{iff}\quad (t_1=t_2\text{ occurs on }\B\ \text{ or }\ t_1\text{ is }t_2)\big].$$

\begin{proposition}\label{prop::Equivalence}
$\sim$ is an equivalence relation.
\end{proposition}


\begin{proposition}\label{prop::Congruence}
	For any $t_1,t_2 \in \mathsf{TERM}(\B)$, if $t_1 \sim t_2$, then $\varphi[x/t_1] \in \B$ iff $\varphi[x/t_2] \in \B$, for all formulas $\varphi$.
\end{proposition}


So equipped, we are ready to prove the cornerstone result of this section.

\begin{lemma}[Satisfaction Lemma]\label{lem::SatisfactionLemma}
Let $\Tab$ be a $\TC_\Log$-tableau, for $\Log \in \{\PFL,\linebreak\PQFL,\NFL,\NQFL,\NQFLm\}$, and let $\B$ be an open and fully expanded branch of $\Tab$. Then there exists a structure $\M_\B^\Log = \langle \D_\B^\Log,{\D_\E}_\B^\Log, \I_\B^\Log\rangle$ and a function $v_\B^\Log$ such that:
\begin{align}\label{cond::Satisfaction}
\text{if}\qquad\psi \in \B,\qquad \text{then}\qquad\M_\B^\Log , v_\B^\Log \models \psi.\tag{$\star$}
\end{align}
\end{lemma}

\begin{proof}
We first show how to construct $\M_\B^\Log$ and $v_\B^\Log$. The latter object is assumed to serve as an assignment, which is normally defined for
 $\mathsf{VAR}(\B) \cup \mathsf{PAR}(\B)$. The values of bound variables, however, are arbitrary, so for convenience we introduce an extra object $\bm{o} \notin \mathsf{TERM}(\B)$ that will further play the role of their value. First we define  $\D_\B^\Log$ and ${\D_\E}_\B^\Log$.

\begin{itemize}
	\item $\D_\B^\Log = \{[t]_\sim \mid t \in \mathsf{TERM}(\B)\}\cup\{\bm{o}\} $.
\end{itemize}

\noindent For $\Log \in \{\PFL,\NFL\}$:
\begin{itemize}
	\item ${\D_\E}_\B^\Log = \{[t]_\sim \in \D_\B^\Log \mid \E t \in \B\}$ [hence $\bm{o}\in \D_\B^\Log \setminus {\D_\E}_\B^\Log$].
\end{itemize}

\medskip

\noindent For $\Log \in \{\PQFL,\NQFL\}$:
\begin{itemize}
	\item ${\D_\E}_\B^\Log = \{[t]_\sim \in \D_\B^\Log \mid \E t \in \B\}\cup\{\bm{o}\}$.
\end{itemize}

\noindent For $\Log \in \{\NQFL^-\}$:
\begin{itemize}
	\item ${\D_\E}_\B^\Log = \{[t]_\sim \in \D_\B^\Log \mid t \in \mathsf{PAR}(\B)\}\cup\{\bm{o}\}$.
\end{itemize}
Next, we define $v_\B^\Log$ as a function mapping elements from $\mathsf{VAR}(\B) \cup \mathsf{PAR}(\B)$ to $ \D_\B^\Log$ for \PFL{} and \NFL, and as a function from $\mathsf{VAR}(\B) \cup \mathsf{PAR}(\B)$ to ${\D_\E}_\B^\Log$ for \PQFL, \NQFL, and \NQFLm. We let
$$v_\B^\Log(t) = \begin{cases}
[t]_\sim,&\text{if $t$ is a parameter},\\
\bm{o},&\text{if $t$ is a bound variable}.
\end{cases}$$

\begin{itemize}
	\item ${\I_\B^\Log}_{v_\B^\Log}(t) = v_\B^\Log(t)$, for each $t \in \mathsf{PAR}(\B) \cup \mathsf{VAR}(\B)$;
	\item ${\I_\B^\Log}_{v_\B^\Log}(\imath x \varphi) = [t]_\sim$ iff
	$\varphi[x/t]\in \B$ and for any $b\in \mathsf{PAR}(\B)$, if $\varphi[x/b]\in \B$, then $t=b\in \B$
	, for each $\imath x \varphi \in \mathsf{TERM}(\B)$ and $t \in \mathsf{PAR}(\B)$;
	\item $\I_\B^\Log(P) = \{\langle {\I_\B^\Log}_{v_\B^\Log}(t_1), \ldots, {\I_\B^\Log}_{v_\B^\Log}(t_n)\rangle \mid P(t_1,\ldots,t_n) \in \B \}$.
\end{itemize}

We need to show that $v_\B^\Log$ is a properly defined $\Log$-assignment.\smallskip

\noindent\textbf{Assignment $v_\B^\Log$}\smallskip
	
\noindent First, we show that $v_\B^\Log$ is a properly defined $\Log$-assignment, for $\Log$ being any of the considered logics. First we prove that $v_\B^\Log$ is a function on $\mathsf{VAR}(\B)$. Totality of $v_\B^\Log$ straightforwardly follows from its definition. Uniqueness of the value assigned by $v_\B^\Log$ to each element of $\mathsf{VAR}(\B)\cup\mathsf{PAR}(\B)$ is a consequence of two facts. First, $\sim$ is an equivalence relation, so equivalence classes of $\sim$ are pairwise disjoint. Secondly, $\D_\B^\Log$ is non-empty. Indeed, without loss of generality we can assume that we check for validity of universally quantified formulas, that is, the input formula $\varphi$ is of the form $\neg \forall x \psi$. By expandedness of $\B$ we get that the rules $(\neg\neg E)$, $(\land E)$, $(\neg\land E)$, $(\neg \forall_i)$, and $(\neg \forall_i)$, for $i\in\{1,2\}$, were applied on $\B$ to the point where an atomic formula or a negated atomic formula with a free term $t$, that is, a parameter or definite description, occurrs on $\B$. Such a formula must finally occur on $\B$ as $\Lan$ does not contain the constants $\bot$ and $\top$ and an atomic formula of $\Lan$ is of one of the forms: $t_1 = t_2$, $P(t_1,\ldots,t_n)$, or $\E t$, where $t, t_1,\ldots,t_n$ are terms and $P$ is an $n$-ary predicate symbol. Thus, an equivalence class of such a freely occurring term $t$ is an element of $\D_\B^\Log$.\smallskip

\noindent For $\Log \in \{\PQFL,\NQFL,\NQFLm\}$ we additionally need to show that the image of $v_\B^\Log$ is included in ${\D_\E}_\B^\Log$. But for the first two logics this is a straightforward consequence of presence of the rule $(\E I_3)$ in $\TC_\PQFL$ and $\TC_\NQFL$, which, for each parameter $b$ on $\B$, introduces $\E b$ to $\B$, and the definition of ${\D_\E}_\B^\Log$ for both logics. In the last case the required inclusion rests solely on the definition of ${\D_\E}_\B^\NQFLm$.\medskip

\noindent Let us now show that (\ref{cond::Satisfaction}) holds. The notion of satisfaction in $\M_\B^\Log$ is defined as in Section~\ref{subsect::Semantics}. We proceed by induction on the complexity of $\psi$ which is defined as the number of connectives and quantifiers occuring in $\psi$ but not in the scope of the $\imath$-operator. We restrict attention to the cases where $\psi:=t_1=t_2$ and $\psi:= t_1\neq t_2$. The proof of the remaining cases can be found in the \hyperref[sect::appendix]{Appendix}.\medskip
 
 \noindent $\psi := t_1 = t_2$\quad Let $t_1,t_2 \in \mathsf{TERM}(\B)$ and $t_1 = t_2 \in \B$. Let $\Log \in \{\PFL,\PQFL\}$. By the definition of $\sim$, $[t_1]_\sim = [t_2]_\sim$, and so, by the definition of ${\I_\B^\Log}_{v_\B^\Log}$, ${\I_\B^\Log}_{v_\B^\Log}(t_1) = {\I_\B^\Log}_{v_\B^\Log}(t_2)$. Thus, by the satisfaction condition for $=$-formulas in both logics, $\M_\B^\Log , v_\B^\Log \models t_1 = t_2$. Now let $\Log \in \{\NFL,\NQFL\}$. By expandedness of $\B$ we know that the rule $(\E I_2)$ (\NFL) or $(\E I_2)$ together with $(\E I_3)$ (\NQFL) was applied to $t_1 = t_2$, thus yielding $\E t_1, \E t_2 \in \B$. By the proof of the case $\psi := \E t$ we know that ${\I_\B^\Log}_{v_\B^\Log} (t_1) \in {\D_\E}_\B^\Log$ and ${\I_\B^\Log}_{v_\B^\Log} (t_2) \in {\D_\E}_\B^\Log$. Moreover, by the definition of $\sim$ and ${\I_\B^\Log}_{v_\B^\Log}$, ${\I_\B^\Log}_{v_\B^\Log} (t_1) = {\I_\B^\Log}_{v_\B^\Log} (t_2)$. Hence, by the satisfaction condition for $=$-formulas, $\M_\B^\Log, v_\B^\Log \models t_1 = t_2$. Finally, let $\Log = \NQFLm$. By expandedness of $\B$ the rule $(= I_2)$ was applied to $t_1 = t_2$, thus yielding $a_i=t_i$, for $1 \leq i \leq 2$ and $t_i$ being a definite description. Without loss of generality assume that $t_1 \in \mathsf{PAR}(\B)$ and $t_2$ is a definite description, so we have $t_1,a_2 \in \mathsf{PAR}(\B)$ and $t_2 = a_2 \in \B$. By the definition of $\sim$ and ${\D_\E}_\B^\Log$ for $\Log = \NQFLm$ we get that $[t_1]_\sim \in {\D_\E}_\B^\Log$, $[t_2]_\sim = [a_2]_\sim \in {\D_\E}_\B^\Log$ and $[t_1]_\sim = [t_2]_\sim$. By the definition of ${\I_\B^\Log}_{v_\B^\Log}$, ${\I_\B^\Log}_{v_\B^\Log}(t_1), {\I_\B^\Log}_{v_\B^\Log}(t_2) \in {\D_\E}_\B^\Log$ and ${\I_\B^\Log}_{v_\B^\Log} (t_1) = {\I_\B^\Log}_{v_\B^\Log} (t_2)$. Hence, by the satisfaction condition for $=$-formulas, $\M_\B^\Log, v_\B^\Log \models t_1 = t_2$.\medskip
 
 \noindent $\psi:= t_1\neq t_2$\quad Let $t_1,t_2 \in \mathsf{TERM}(\B)$ and $t_1 \neq t_2 \in \B$. Let $\Log \in \{\PFL,\PQFL\}$. By openness of $\B$, $t_1$ and $t_2$ are distinct terms, for otherwise the rule $(\bot_2)$ would close $\B$. Again, by openness of $\B$, $t_1 = t_2 \notin \B$, so by the definition of $\sim$, $[t_1]_\sim \neq [t_2]_\sim$. Hence, by the definition of ${\I_\B^\Log}_{v_\B^\Log}$, ${\I_\B^\Log}_{v_\B^\Log}(t_1) \neq {\I_\B^\Log}_{v_\B^\Log}(t_2)$. Thus, by the satisfaction condition for $=$-formulas in both logics, $\M_\B^\Log , v_\B^\Log \not\models t_1 = t_2$, and so, by the satisfaction condition for $\neg$-formulas, $\M_\B^\Log , v_\B^\Log \models t_1 \neq t_2$. Let $\Log \in \{\NFL\}$. Clearly, either $t_1$ and $t_2$ are distinct, or identical. Assume, first, that $t_1$ and $t_2$ are distinct terms. Then we proceed with the proof similarly to the case for $\Log \in \{\PFL, \PQFL\}$. Now, assume that $t_1 \neq t_2$ is of one of the forms $t \neq t$. We know that $\E t \notin \B$, for otherwise we could apply $(\E E_2)$ and close $\B$ with $(\bot_1)$. Then, by the definition of $\sim$ and ${\D_\E}_\B^\Log$, it follows that $[t]_\sim \notin {\D_\E}_\B^\Log$. By the definition of ${\I_\B^\Log}_{v_\B^\Log}$ and the satisfaction condition for $=$-formulas, we get $\M_\B^\Log , v_\B^\Log \not\models t=t$. By the satisfaction condition for $\neg$-formulas we finally obtain $\M_\B^\Log , v_\B^\Log \models t \neq t$. 
 Let $\Log\in\{\NQFL,\NQFLm\}$. Clearly, either $t_1$ and $t_2$ are distinct, or $t_1,t_2\notin\mathsf{PAR}(\B)$. Indeed, if $t_1 \neq t_2$ was of the form $b \neq b$ for $b \in \mathsf{PAR}(\B)$, then $\B$ would be closed by an application of $(\bot_3)$. Assume, first, that $t_1$ and $t_2$ are distinct terms. Then we proceed with the proof similarly to the case for $\Log \in \{\PFL, \PQFL\}$. Now, assume that $t_1 \neq t_2$ is of the form $\imath x \varphi \neq \imath x \varphi$. Let $\Log = \NQFL$. Certainly, $\E \imath x \varphi \notin \B$, for otherwise $(\E E_1)$ would have been applied, yielding $a = \imath x \varphi$ and, through $(=E)$, $a\neq a$, thus closing $\B$. So, by the definition of $\sim$, ${\D_\E}_\B^\Log$, and ${\I_\B^\Log}_{v_\B^\Log}$, we have ${\I_\B^\Log}_{v_\B^\Log}(\imath x \varphi) \notin {\D_\E}_\B^\Log$. The rest of the proof is identical to the one for $\Log = \NFL$. Let $\Log = \NQFLm$. For the same reasons as for $\NQFL$, for each $b \in \mathsf{PAR}(\B)$, $b = \imath x \varphi \notin \B$. Then, by the definition of $\sim$, ${\D_\E}_\B^\Log$, and ${\I_\B^\Log}_{v_\B^\Log}$, ${\I_\B^\Log}_{v_\B^\Log}(\imath x \varphi)\notin {\D_\E}_\B^\Log$. We conduct the rest of the proof similarly to the one for $\Log \in \{\NFL,\NQFL\}$.\medskip

\noindent \textbf{Interpretation ${\I_\B^\Log}_{v_\B^\Log}(\imath x \varphi)$}\medskip

\noindent The last thing we must show is that the  condition for the interpretation of definite descriptions holds in $\M_\B^\Log$. In terms of the induced model it amounts to the following condition:
\begin{align}\label{form::Equivalence}
{\I_\B^\Log}_{v_\B^\Log}(\imath x \varphi) = [a]_\sim\in {\D_\E}_\B^\Log\quad \text{iff}\quad \begin{minipage}[t]{6.4cm} $\M_\B^\Log , {v^x_{[a]_\sim}}_\B^\Log \models \varphi$ and for each $x$-variant ${v'}_\B^\Log$ of $v_\B^\Log$, if $\M_\B^\Log , {v'}_\B^\Log \models \varphi$, then  ${v'}_\B^\Log(x)= [a]_\sim$.
\end{minipage} \tag{$\dagger$}
\end{align}
The right-hand side of (\ref{form::Equivalence}), by  \href{lem::Substitution}{Substitution Lemma}, is equivalent to the condition that $\M_\B^\Log , v_\B^\Log \models \varphi[x/a]$ and for each $b$ such that $[b]_\sim\in {\D_\E}_\B^\Log$, if $\M_\B^\Log , v_\B^\Log \models \varphi[x/b]$, then  $[b]_\sim = [a]_\sim$, which will be applied in the proof. We show (\ref{form::Equivalence}) for \PQFL, \NQFL, \NQFLm. For the remaining systems the proof is similar. First let us note the following:
\begin{claim}\label{prop::DefiniteDescription}
	Let $\B$ be a fully expanded branch of a $\TC_\Log$-tableau $\Tab$, for $\Log \in \{ \PFL, \linebreak\PQFL, \NFL, \NQFL, \NQFLm\}$. Then the following holds:	
		$${\I_\B^\Log}_{v_\B^\Log}(\imath x \varphi) = [a]_\sim\text{ iff }\imath x\varphi=a\in \B.$$
\end{claim}
\begin{proof}
	$\Rightarrow$\quad By contraposition, assume that $\imath x\varphi=a\notin \B$. Then, by $(cut_1)$, $\imath x\varphi\neq a\in \B$, which, by
	(\ref{cond::Satisfaction}), yields that $\M_\B^\Log , v_\B^\Log \models \imath x \varphi\neq a$. Thus, ${\I_\B^\Log}_{v_\B^\Log}(\imath x \varphi) \neq [a]_\sim$.\medskip
	
	\noindent $\Leftarrow$\quad  If we assume that $\imath x\varphi=a\in \B$, then, by (\ref{cond::Satisfaction}), $\M_\B^\Log , v_\B^\Log \models \imath x \varphi = a$, and we are done.
\end{proof}
\noindent Now let us prove (\ref{form::Equivalence}):

\medskip

\noindent $\Rightarrow$\quad Let ${\I_\B^\Log}_{v_\B^\Log}(\imath x \varphi) = [a]_\sim\in {\D_\E}_\B^\Log$. Hence, by \hyperref[prop::DefiniteDescription]{Claim}, $\imath x \varphi= a\in \B$. By $(\imath E_1)$, either $\varphi[x/a] \in \B$ and  $\neg\varphi[x/b] \in \B$, or $\varphi[x/a] \in \B$ and $a=b\in \B$, for every $b$. In both cases, by (\ref{cond::Satisfaction}), $\M_\B^\Log , v_\B^\Log \models \varphi[x/a]$. Moreover, again by (\ref{cond::Satisfaction}), either $\M_\B^\Log , v_\B^\Log \not \models \varphi[x/b]$ or $\M_\B^\Log , v_\B^\Log \models a=b$, for every $b$. Hence the second conjunct follows.

\medskip

\noindent $\Leftarrow$\quad 
Assume that $\M_\B^\Log , v_\B^\Log \models \varphi[x/a]$ and for each $b\in {\D_\E}_\B^\Log$, if $\M_\B^\Log , v_\B^\Log \models \varphi[x/b]$, then  $[b]_\sim = [a]_\sim$, but ${\I_\B^\Log}_{v_\B^\Log}(\imath x \varphi) \neq [a]_\sim$. Hence, by \hyperref[prop::DefiniteDescription]{Claim},
 $\imath x \varphi\neq a\in \B$. 
 By $(\neg\imath E)$ we have either $\neg\varphi[x/a] \in \B$ or, for some $b$, $\varphi[x/b] \in \B$ and $a\neq b \in \B$. Both cases, by 
(\ref{cond::Satisfaction}), lead to a contradiction.
\end{proof}

\begin{theorem}[Completeness]\label{thm::Completeness}
	The tableau calculi $\TC_\PFL$, $\TC_\PQFL$, $\TC_\NFL$, $\TC_\NQFL$, and $\TC_\NQFLm$ are complete.	
\end{theorem}

\begin{proof}
	We prove the contrapositive of the usual completeness condition. Assume that a $\Log$-formula $\varphi$ is not tableau-valid wrt $\TC_\Log$. Then, there is a fully expanded $\TC_\Log$-tableau which is not a tableau proof of $\varphi$. Thus, there exists an open branch $\B$ in $\T$ with $\neg \varphi$ at the root. By \href{lem::SatisfactionLemma}{Satisfaction Lemma} the structure $\M_\B^\Log = \langle \D_\B^\Log, {\D_\E}_\B^\Log, \I_\B^\Log\rangle$ is an $\Log$-model and the function $v_\B^\Log: \mathsf{VAR}\cup \mathsf{PAR}(\B) \longrightarrow \D_\B^\Log$ is an $\L$-assignment and since $\neg \varphi \in \B$, then $\M_\B^\Log , v_\B^\Log \models \neg \varphi$. By the usual duality between satisfiability and validity we obtain that $\varphi$ is not valid, which yields the conclusion.
\end{proof}

\section{Related Work}\label{sect::Related}

Alongside with the tableau systems mentioned in Section~\ref{sect::Introduction}, which usually directly transform the conditions (\ref{cond::Lambert}) or (\ref{cond::Russell}), two alternative approaches deserve a separate mention. One of them, although in the setting of labelled sequent calculus, has recently been presented by Orlandelli~\cite{orlandelli}. He provided an alternative formulation of modal theory of descriptions developed by Fitting and Mendelsohn in~\cite{fitt:fir98} in the form of a tableau system not enjoying the subformula property. Orlandelli's system, on the other hand, is cut-free and analytic. These properties are obtained at the cost of a significant enrichment of the technical machinery. In addition to ordinary strong labels (i.e., labels naming worlds and attached to formulas and relational atoms showing accessibility links between worlds), he is using special denotation atoms $D(t,x,w)$ to express that a term $t$ in $w$ denotes the same object as the one denoted by a variable $x$. This device is used to define rules for DD and  for the $\lambda$-operator.
Another cut-free formulation of the same theory of descriptions was developed by Indrzejczak~\cite{ind:aml20} in the setting of hybrid modal language. The main difference is that instead of introducing external labelling apparatus, a richer language with nominal variables and sat-operators is used and descriptions are characterized by means of rules dealing with equalities, like in the present approach.
\textsf{MFD} in all variants analyzed in the present paper is a much weaker theory of descriptions than the theory mentioned above, although the variants based on \NQFL{} and \NQFLm{} show some affinities with Fitting and Mendelsohn's theory. It would be an interesting task to embed \textsf{MFD}, as represented in positive free logic, in the modal setting using one of the two presented alternative approaches.

The tableau calculi devised in this paper, despite being based on the cut-free sequent calculi for the same logics, introduced in~\cite{ind:llp20}, go beyond straightforward transpositions of the rules presented therein. The main aim of~\cite{ind:llp20} was to obtain sequent formalizations of free logics for which it is possible to prove the cut elimination theorem in a constructive way. Our main objective here is to construct calculi which are analytic and effective tools of proof search in respective logics. This basic difference has a significant impact on the way the sets of rules are built in both approaches, which we briefly summarize in what follows.
First of all, in our tableau systems a restricted (to identities) form of analytic cut is present, whereas in the sequent calculus from~\cite{ind:llp20} cut is in general constructively eliminable. However, cut-freeness of the latter systems leads to more complicated forms of some other rules. In particular:
\begin{enumerate}
\item The sequent counterpart of the tableau rule $(=E)$ is restricted to atomic formulas and has three premises instead of one.
\item Some sequent rules are replaced here by suitable closure rules.
\item All tableau rules for definite descriptions are different than the respective rules in sequent calculi.
\end{enumerate}
What speaks in favour of tableaux presented in this paper is a decreased branching factor in comparison to the discussed sequent calculi. The price to be paid, however, is a restricted form of analytic cut which is necessary to ensure completeness of the calculi. Since eliminating the three-premise rule makes it necessary to add a resticted cut, we cannot be sure that it leads to simpler proof-trees in the general case, but, at least on the basis of several tested examples, it seems highly probable.

The presence of cut, even in a strictly limited form which does not destroy the subformula property, may be seen as a disadvantage. However,
both cut rules could be dispensed with and replaced with two other rules expressing some form of Leibniz's law:
	$$(RL_1)\ \dfrac{\neg\varphi[x/t]}{\neg\varphi[x/b] \mid b \neq t}\qquad
	(RL_2)\ \dfrac{\neg\varphi[x/t], \E b}{\neg\varphi[x/b] \mid b \neq t},$$
	where $t$ is a DD and $\varphi$ is atomic (including $\E$ and $=$). 
On the other hand, in comparison to the above Leibniz's rules the proposed form of analytic cut seems to be a more direct solution without overhead costs. The cut-free and analytic characterization of Russellian theory of DD from~\cite{ind:des21} is essentially based on the introduction of a collection of special equlity rules for every kind of involved terms. Only after we augment the calculus with this extra toolkit, it becomes possible to dispense with any form of cut. However, despite of some purely proof-theoretic advantages of this solution, it does not seem to bring any serious benefits in the tableau setting.
\section{Conclusions}\label{sect::Conclusions}

The role of definite descriptions in the field of proof theory and automated deduction has so far been underestimated. That is why it is important to stress advantages using them may bring. First of all, as we mentioned in Section~\ref{sect::Preliminaries}, every complex term represented by means of functional terms can be equivalently expressed using a definite description. In the latter case we do not need extra bridge principles showing how the information encoded by functional terms is represented by predicates, whereas in the former case we do. For example such bridge principles are usually needed as enthymematic premises in an analysis of obviously valid arguments. Moreover, the presence of functions in formal languages often easily leads to generating infinite Herbrand models even when finite models are allowed. Let us illustrate this with a simple example. From $\forall x(a=f(x))$ we infer $a=f(a), a=f(f(a)), a=f(f(f(a))), \ldots$ On the other hand, from $\forall x(a=\imath y F(x,y))$ we obtain $a=\imath yF(a,y)$, and then $F(a,a), \neg F(a,a) \ | \ a = a, F(a,a)$, where the left branch gets closed, but the right one provides a finite, single-element model. Moreover, definite descriptions can be used to provide smooth definitions of new terms, and even new operators, in formal languages. For example, one may define the abstraction operator in set theory in an elegant way.

These virtues of definite descriptions have not hitherto been thoroughly examined mainly because of a lack of good formal systems expressing their theories. The presented tableau systems are a step towards filling this gap. They are analytic despite of the use of restricted cuts and, in effect, seem to provide handy proof-search tools. Further plans for research include:

\begin{enumerate}
	\item designing and implementing a tool for automated proof-search and user-friendly proof-assistance;
	\item investigating computational efficiency of such a tool; in particular, comparing it with well-known programs designed for standard languages with functional terms;
	\item formalizing stronger theories of definite descriptions in standard language and in enriched languages (e.g., with modalities);
	\item applying these systems to a formalization of elementary theories.
\end{enumerate}

\bibliographystyle{splncs04}
\bibliography{biblio}

\newpage

\begin{subappendices}
	\renewcommand{\thesection}{\Alph{section}}
	\section{Omitted proofs}\label{sect::appendix}

\subsection*{Proof of symmetry of $=$}

\begin{proof}
	In all tableau systems the $=$ relation is symmetric, that is, if a formula $t_1=t_2$ is derivable on a branch $\mathscr{B}$, then a formula $t_2=t_1$ is derivable on $\mathscr{B}$, too:
	\begin{center}
		\def\arraystretch{1.2}
		
	\begin{tabular}{lll}
		& \qquad\quad $\vdots$ &\\
	$n$ & \qquad $t_1 = t_2$ &\\
$n+1$ & \qquad $t_1 = t_1$ & \qquad $(= E)$: $n$ (twice)\\
$n+2$ & \qquad $t_2 = t_1$& \qquad $(= E)$: $n$, $n+1$\\
	& \qquad\quad $\vdots$ &
	\end{tabular}

\qedhere
\end{center}
\end{proof}

\subsection*{Proof of \Cref{lem::SoundRules}}

\begin{replemma}{lem::SoundRules}
	For each $\Log \in \{\PFL, \PQFL, \NFL, \NQFL, \NQFLm\}$ all rules of $\TC_\Log$ are sound.
\end{replemma}

\begin{proof}
	Soundness of $(\neg\neg E)$, $(\land E)$, and $({\neg}{\land} E)$ straightforwardly follows from the satisfaction conditions from Section~\ref{subsect::Semantics}.\smallskip
	
	\noindent$(\bot_1)$, $(\bot_2)$, and $(\bot_3)$ are ``vacuously'' sound, as in all three cases the premises are unsatisfiable in respective logics (which also is a direct consequence of the satisfaction conditions from Section~\ref{subsect::Semantics}).\smallskip
	
	\noindent$(\forall E_1)$ and $(\neg\forall E_1)$ are standard rules for quantifier elimination without any constraints on parameters replacing the variables bounded by the quantifier. Assume that $\forall x \varphi$ is $\Log$-satisfiable, where $\Log \in \{\PQFL, \NQFL, \NQFLm\}$. Then, there exists a model $\M = \langle \D, \D_\E, \I\rangle$ and an assignment $v$ such that for any object $o \in \D_\E$, $\M, v_o^x \models \varphi$. Let $b$ be a variable present on the branch and let $v(b) = o'$. $\Log \in \{\PQFL,\NQFL,\NQFLm\}$, so we know that $o' \in \D_\E$. Then, by \href{lem::Substitution}{Substitution Lemma} we get that $\M , v \models \varphi[x/b]$. Now, assume that $\neg\forall x \varphi$ is $\Log$-satisfiable, where $\Log \in \{\PQFL, \NQFL, \NQFLm\}$. Then, there exists a model $\M = \langle \D, \D_\E, \I\rangle$, an assignment $v$, and an object $o\in\D_\E$ such that $\M, v_o^x \models \neg \varphi$. Let now $a$ be a fresh variable not-occurring in $\varphi$ and let $v'$ be an assignment such that $v'(y) = v(y)$, for each variable $y$ occurring in $\varphi$, and $v'(a) = o = v'(x)$. Then by \href{lem::Coincidence}{Coincidence Lemma} we get $\M, {v}_{o}^x \models \neg\varphi$ and by \href{lem::Substitution}{Substitution Lemma} we finally obtain $\M, v \models \neg\varphi[x/a]$.\smallskip
	
	\noindent The proof for $(\forall E_2)$ and $(\neg\forall E_2)$ occuring in $\TC_\Log$ for $\Log \in \{\PFL, \NFL\}$ is analogical with the one above. The only proviso that has to be carried is that since parameters may be non-denoting in both considered logics, we need to externally guarantee that all parameters involved in the rules are denoting. We do so by putting them in the scope of the existence predicate $\E$.\smallskip
	
	\noindent In the remainder of the proof we will use both sides of the equivalence in \href{lem::Substitution}{Substitution Lemma} interchangeably without explicitly mentioning that.\smallskip
	
	Soundness of $(= E)$ can be proven by an induction on the complexity of $\varphi$. We show the base case for an atomic formula of the form $P(t_1,\ldots,t_n)$, and leave the remaining (base) cases to the reader. Assume that $t_1 = t_2, P(t_1,t'_2\ldots,t'_n)$ are $\Log$-satisfiable, for $\Log \in \{\PFL, \PQFL, \NFL, \NQFL, \NQFLm\}$. Then there exists a model $\M = \langle \D, \D_\E, \I\rangle$ and an assignment $v$ such that $\M, v \models t_1 = t_2$ and $\M, v \models P(t_1,t'_2,\ldots,t'_n)$. Then $\langle \I_v(t_1),\I_v(t'_2),\ldots,\I_v(t'_n)\rangle \in \I(P)$ and $\I_v(t_1) = \I_v(t_2)$. Therefore, $\langle \I_v(t_2),\I_v(t'_2),\ldots,\I_v(t'_n)\rangle \in \I_v(P)$, which finally yields $\M, v \models P(t_2,t'_2,\ldots,t'_n)$. Now for the inductive step suppose that the rule is sound for $\varphi$. We will show that it is sound for $\forall x \varphi$. Assume that $t_1 = t_2, \forall x \varphi$ are $\Log$-satisfiable, that is, there exists a model $\M = \langle \D, \D_\E, \I\rangle$ and an assignment $v$ such that $\M, v \models t_1 = t_2$ and $\M, v \models \forall x \varphi$. If $t_1 = x$, then $t_1$ does not occur freely in $\forall x \varphi$, and so, $\M, v \models (\forall x \varphi)[t_1 / t_2]$ since $(\forall x \varphi)[t_1 / t_2] = \forall x \varphi$. Assume, then, that $x \neq t_1$. It means that for each variable $b$ such that $v(b) \in \D_\E$, $\M, v \models \varphi[x/b]$. By the inductive assumption, $\M, v \models \varphi[x/b,t_1/t_2]$. By the arbitrariness of $b$ we get $\M, v \models \forall x\varphi[t_1 / t_2]$. The remaining inductive steps are left to the reader.\smallskip
	
	\noindent $(= I_1)$ and $(= I_2)$ occur only in $\TC_\NQFLm$ which does not feature the existence predicate $\E$. Assume that $P(t_1,\ldots,t_n)$ is \NQFLm-satisfiable. It means that there exists a model $\M = \langle \D,\D_\E,\I\rangle$ and an assignment $v$ such that $\M,v \models P(t_1, \ldots, t_n)$. Hence, $\langle \I_v(t_1), \ldots, \I_v(t_n)\rangle \in \I(P)$. Let $t_i$ be a definite description. Thus, there exists a variable $x$ and an \NQFLm-formula $\varphi$ such that $v(x) = o \in \D_\E$, $\M , v \models \varphi(x)$ and for any variable $y$, $\M , v \models \varphi(y)$ iff $x = y$. Therefore, $\I_v(t_i) = o$. Let $a_i$ be a variable not occurring among $t_1, \ldots, t_n$ and not occurring in $\varphi$. We extend $v$ to $v'$ by setting $v'(a_i) = o$. Since $v(a_i) = v(x) = o$, $\I_v(t_i) = o$, and so, $\M , v' \models P(t_1, \ldots, t_n), a_i = t_i$ as required. Proving soundness of $( = I_2)$ is conducted similarly.\smallskip
	
	\noindent Soundness of $(cut_1)$ and $(cut_2)$ is a direct consequence of them being a restricted form of analytic cut, so an meta-instance of the law of excluded middle.\smallskip
	
	\noindent In order to prove soundness of $(\E E_1)$ assume that $\E t$, where $t$ is a definite description, is $\Log$-satisfiable, for $\Log \in \{\PFL,\PQFL,\NFL,\NQFL\}$, that is, there exists a model $\M = \langle \D, \D_\E, \I\rangle$ and an assignment $v$ such that $\M , v \models \E t$. By the condition for satisfaction of $\E$-formulas we know that $\I_v(t) \in \D_\E$. Let $\varphi$ be a formula occurring in $t$. There exists a variable $x$ such that $v(x) = \I_v(t) = o \in \D_\E$, $\M, v \models \varphi(x)$ and for any variable $y$, $\M , v \models \varphi(y)$ iff $x = y$. Let $a$ be a variable not occurring in $\varphi$. We extend $v$ to $v'$ by setting $v'(a) = o$. Then $\I_v(a) = \I_v(x)$, and so, $\I_v(a) = \I_v(t)$. Hence, $\M ,v' \models \E t, a=t$. Soundness of $(\E E_2)$ straightforwardly follows from reflexivity of equality occurring on the right-hand side of the satisfaction condition for $=$-formulas.\smallskip
	
	
	\noindent To show that $(\E I_1)$ is sound, assume that $P(t_1,\ldots,t_n)$ is \NQFL-satisfiable. It means that there exists a model $\M = \langle \D, \D_\E, \I\rangle$ and an assignment $v$ such that $\M, v \models P(t_1,\ldots,t_n)$. Hence, $\langle \I_v(t_1),\ldots,\I_v(t_n)\rangle \in \I(P)$ and $\I_v(t_i) \in \D_\E$, for $1 \leq i \leq n$ and $t_i$ being a definite description. Thus, by the satisfaction condition for $\E$-formulas we get that $\M, v \models \E t_i$, as required. A proof of soundness of $(\E I_2)$ is analogical. To prove that $(\E I_3)$ is sound, it suffices to recall that all variables in \PQFL{} and \NQFL{} are denoting. Soundness of $(\E I_4)$ is a straightforward consequence of the assumption that we consider only models with non-empty domains $\D_\E$ of existing objects.\smallskip

\noindent To prove soundness of $(\imath E_1)$ assume that $b_1 = \imath x \varphi$ is $\Log$-satisfiable, for $\Log \in \{\PQFL, \NQFL, \NQFLm\}$, that is, there exists a model $\M = \langle \D, \D_\E, \I \rangle$ and an assignment $v$ such that $\M, v \models b_1 = \imath x \varphi$. Let $v(b_1)= o\in \D_\E$, then $\I_v(\imath x \varphi) = v(b_1) = o$ and by the satisfaction condition $\M, v^x_{o} \models \varphi$, and for any $x$-variant $v'$ of $v$, if $\M, v' \models \varphi$, then $v'(x)=o$. The first conjunct guarantees, by \href{lem::Substitution}{Substitution Lemma}, that $\M, v \models \varphi[x/b_1]$, which holds for both conclusions. The second conjunct yields, for any $b_2\in \D_\E$, that either $\M , v \not \models \varphi[x/b_2]$ or $\M , v \models b_1 = b_2$. The former case yields the left conclusion, whereas the latter case yields the right one.
To show that $(\neg \imath E_1)$ is sound assume that $b \neq \imath x  \varphi$ is $\Log$-satisfiable for $\Log \in \{\PQFL, \NQFL, \NQFLm\}$. Then, there exists a model $\M = \langle \D, \D_\E, \I \rangle$ and an assignment $v$ such that $\M , v \models b \neq \imath x \varphi$. It means that $\I_v(\imath x \varphi) \neq v(b)=o \in \D_\E$. By the satisfaction condition $\M, v^x_{o} \not \models \varphi$, or for some $x$-variant $v'$ of $v$, $\M, v' \models \varphi$ but $o'=v'(x)\neq v(x)=o$. In the first case, by \href{lem::Substitution}{Substitution Lemma}, $\M, v \not \models \varphi[x/b]$, so the left conclusion is satisfied. If the second holds, then by \href{lem::Coincidence}{Coincidence Lemma} and \href{lem::Substitution}{Substitution Lemma} we have that $\M, v \models \varphi[x/a]$ but  $\M , v \models b \neq a$ for some fresh $a$.\smallskip

\noindent Proofs for $(\imath E_2)$ and $(\neg \imath E_2)$, respectively, are conducted analogically with the following caveat. In \PFL{} and \NFL{} variables are not automatically guaranteed to denote, so the existence of a referrent object needs to be ensured externally. This is done by placing a variable in the scope of the existence predicate~$\E$.
\end{proof}

\subsection*{Proof of \Cref{prop::Equivalence}}

\begin{repproposition}{prop::Equivalence}
	$\sim$ is an equivalence relation.
\end{repproposition}

\begin{proof}
	Indeed, \emph{reflexivity} of $\sim$ follows from the second disjunct occurring on the right-hand side of the above equivalence. \emph{Symmetry} is ensured by the fact that in $\TC_\Log$ the $=$ relation is symmetric, that is, if $t_1=t_2$ occurs on $\B$, so does $t_2=t_1$. For \emph{transitivity} assume that for $t_1,t_2,t_3 \in \mathsf{TERM}(\B)$, $t_1\sim t_2$ and $t_2\sim t_3$. If $t_1$ is identical to $t_2$ or $t_2$ is identical to $t_3$, then we straightforwardly get $t_1\sim t_3$. Assume that $t_1$ is distinct than $t_2$ and $t_2$ is distinct than $t_3$. Then, $t_1=t_2$ and $t_2=t_3$ occurred on $\B$. Since $\B$ is fully expanded, $(= E)$ was applied to $t_1=t_2$ and $t_2=t_3$, thus introducing $t_1=t_3$ to $\B$. Hence, $t_1\sim t_3$.
\end{proof}

\subsection*{Proof of \Cref{prop::Congruence}}

\begin{repproposition}{prop::Congruence}
	For any $t_1,t_2 \in \mathsf{TERM}(\B)$, if $t_1 \sim t_2$, then $\varphi[x/t_1] \in \B$ iff $\varphi[x/t_2] \in \B$, for all formulas $\varphi$.
\end{repproposition}

\begin{proof}
	Let $t_1,t_2 \in \mathsf{TERM}(\B)$ be such that $t_1 \sim t_2$. Then either $t_1 = t_2 \in \B$, or $t_1$ is identical to $t_2$. In the latter case the claim follows trivially. In the former case assume that $t_1 = t_2, \varphi[x/t_1] \in \B$. Then by an application of the rule $(=E)$ we obtain that $\varphi[x/t_2] \in \B$. The proof of the reverse implication is analogous.
\end{proof}

\subsection*{Proof of the inductive part of \Cref{lem::SatisfactionLemma}}

\begin{replemma}{lem::SatisfactionLemma}[\textbf{Satisfaction Lemma}]
	Let $\Tab$ be a $\TC_\Log$-tableau, for $\Log \in \{\PFL,\linebreak\PQFL,\NFL,\NQFL,\NQFLm\}$, and let $\B$ be an open and fully expanded branch of $\Tab$. Then there exists a structure $\M_\B^\Log = \langle \D_\B^\Log,{\D_\E}_\B^\Log, \I_\B^\Log\rangle$ and a function $v_\B^\Log$ such that:
	\begin{align}\label{cond::Satisfaction1}
		\text{if}\qquad\psi \in \B,\qquad \text{then}\qquad\M_\B^\Log , v_\B^\Log \models \psi.\tag{$\star$}
	\end{align}
\end{replemma}

\begin{proof}
 Let us now show that (\ref{cond::Satisfaction1}) holds. The notion of satisfaction in $\M_\B^\Log$ is defined as in Section~\ref{subsect::Semantics}. We proceed by induction on the complexity of $\psi$ which is defined as the number of connectives and quantifiers occuring in $\psi$ but not in the scope of the $\imath$-operator.\medskip
	
	\noindent $\psi := \E t$\quad Let $t \in \mathsf{TERM}(\B)$, $\E t \in \B$, and $\Log \in \{\PFL, \PQFL, \NFL, \NQFL\}$. By the construction of ${\D_\E}_\B^\Log$ and the definition of ${\I_\B^\Log}_{v_\B^\Log}$ we get that ${\I_\B^\Log}_{v_\B^\Log} (t) = [t]_\sim \in {\D_\E}_\B^\Log$, and so, $\M_\B^\Log , v_\B^\Log \models \E(t)$.\medskip
	
	\noindent $\psi := \neg \E t$\quad Let $t \in \mathsf{TERM}(\B)$, $\E t \in \B$, and $\Log \in \{\PFL, \PQFL, \NFL, \NQFL\}$. By openness of $\B$ we get that $\E t \notin \B$. Therefore, by the construction of ${\D_\E}_\B^\Log$ and the definition of ${\I_\B^\Log}_{v_\B^\Log}$ we get that ${\I_\B^\Log}_{v_\B^\Log} (t) = [t]_\sim \notin {\D_\E}_\B^\Log$, and so, $\M_\B^\Log , v_\B^\Log \not\models \E(t)$. Hence, by the satisfaction condition for $\neg$-formulas, $\M_\B^\Log , v_\B^\Log \models \neg \E(t)$.\medskip
	
	 \noindent $\psi := t_1 = t_2$\quad Let $t_1,t_2 \in \mathsf{TERM}(\B)$ and $t_1 = t_2 \in \B$. Let $\Log \in \{\PFL,\PQFL\}$. By the definition of $\sim$, $[t_1]_\sim = [t_2]_\sim$, and so, by the definition of ${\I_\B^\Log}_{v_\B^\Log}$, ${\I_\B^\Log}_{v_\B^\Log}(t_1) = {\I_\B^\Log}_{v_\B^\Log}(t_2)$. Thus, by the satisfaction condition for $=$-formulas in both logics, $\M_\B^\Log , v_\B^\Log \models t_1 = t_2$. Now let $\Log \in \{\NFL,\NQFL\}$. By expandedness of $\B$ we know that the rule $(\E I_2)$ (\NFL) or $(\E I_2)$ together with $(\E I_3)$ (\NQFL) was applied to $t_1 = t_2$, thus yielding $\E t_1, \E t_2 \in \B$. By the proof of the case $\psi := \E t$ we know that ${\I_\B^\Log}_{v_\B^\Log} (t_1) \in {\D_\E}_\B^\Log$ and ${\I_\B^\Log}_{v_\B^\Log} (t_2) \in {\D_\E}_\B^\Log$. Moreover, by the definition of $\sim$ and ${\I_\B^\Log}_{v_\B^\Log}$, ${\I_\B^\Log}_{v_\B^\Log} (t_1) = {\I_\B^\Log}_{v_\B^\Log} (t_2)$. Hence, by the satisfaction condition for $=$-formulas, $\M_\B^\Log, v_\B^\Log \models t_1 = t_2$. Finally, let $\Log = \NQFLm$. By expandedness of $\B$ the rule $(= I_2)$ was applied to $t_1 = t_2$, thus yielding $a_i=t_i$, for $1 \leq i \leq 2$ and $t_i$ being a definite description. Without loss of generality assume that $t_1 \in \mathsf{PAR}(\B)$ and $t_2$ is a definite description, so we have $t_1,a_2 \in \mathsf{PAR}(\B)$ and $t_2 = a_2 \in \B$. By the definition of $\sim$ and ${\D_\E}_\B^\Log$ for $\Log = \NQFLm$ we get that $[t_1]_\sim \in {\D_\E}_\B^\Log$, $[t_2]_\sim = [a_2]_\sim \in {\D_\E}_\B^\Log$ and $[t_1]_\sim = [t_2]_\sim$. By the definition of ${\I_\B^\Log}_{v_\B^\Log}$, ${\I_\B^\Log}_{v_\B^\Log}(t_1), {\I_\B^\Log}_{v_\B^\Log}(t_2) \in {\D_\E}_\B^\Log$ and ${\I_\B^\Log}_{v_\B^\Log} (t_1) = {\I_\B^\Log}_{v_\B^\Log} (t_2)$. Hence, by the satisfaction condition for $=$-formulas, $\M_\B^\Log, v_\B^\Log \models t_1 = t_2$.\medskip
	
	\noindent $\psi:= t_1\neq t_2$\quad Let $t_1,t_2 \in \mathsf{TERM}(\B)$ and $t_1 \neq t_2 \in \B$. Let $\Log \in \{\PFL,\PQFL\}$. By openness of $\B$, $t_1$ and $t_2$ are distinct terms, for otherwise the rule $(\bot_2)$ would close $\B$. Again, by openness of $\B$, $t_1 = t_2 \notin \B$, so by the definition of $\sim$, $[t_1]_\sim \neq [t_2]_\sim$. Hence, by the definition of ${\I_\B^\Log}_{v_\B^\Log}$, ${\I_\B^\Log}_{v_\B^\Log}(t_1) \neq {\I_\B^\Log}_{v_\B^\Log}(t_2)$. Thus, by the satisfaction condition for $=$-formulas in both logics, $\M_\B^\Log , v_\B^\Log \not\models t_1 = t_2$, and so, by the satisfaction condition for $\neg$-formulas, $\M_\B^\Log , v_\B^\Log \models t_1 \neq t_2$. Let $\Log \in \{\NFL\}$. Clearly, either $t_1$ and $t_2$ are distinct, or identical. Assume, first, that $t_1$ and $t_2$ are distinct terms. Then we proceed with the proof similarly to the case for $\Log \in \{\PFL, \PQFL\}$. Now, assume that $t_1 \neq t_2$ is of one of the forms $t \neq t$. We know that $\E t \notin \B$, for otherwise we could apply $(\E E_2)$ and close $\B$ with $(\bot_1)$. Then, by the definition of $\sim$ and ${\D_\E}_\B^\Log$, it follows that $[t]_\sim \notin {\D_\E}_\B^\Log$. By the definition of ${\I_\B^\Log}_{v_\B^\Log}$ and the satisfaction condition for $=$-formulas, we get $\M_\B^\Log , v_\B^\Log \not\models t=t$. By the satisfaction condition for $\neg$-formulas we finally obtain $\M_\B^\Log , v_\B^\Log \models t \neq t$. 
	Let $\Log\in\{\NQFL,\NQFLm\}$. Clearly, either $t_1$ and $t_2$ are distinct, or $t_1,t_2\notin\mathsf{PAR}(\B)$. Indeed, if $t_1 \neq t_2$ was of the form $b \neq b$ for $b \in \mathsf{PAR}(\B)$, then $\B$ would be closed by an application of $(\bot_3)$. Assume, first, that $t_1$ and $t_2$ are distinct terms. Then we proceed with the proof similarly to the case for $\Log \in \{\PFL, \PQFL\}$. Now, assume that $t_1 \neq t_2$ is of the form $\imath x \varphi \neq \imath x \varphi$. Let $\Log = \NQFL$. Certainly, $\E \imath x \varphi \notin \B$, for otherwise $(\E E_1)$ would have been applied, yielding $a = \imath x \varphi$ and, through $(=E)$, $a\neq a$, thus closing $\B$. So, by the definition of $\sim$, ${\D_\E}_\B^\Log$, and ${\I_\B^\Log}_{v_\B^\Log}$, we have ${\I_\B^\Log}_{v_\B^\Log}(\imath x \varphi) \notin {\D_\E}_\B^\Log$. The rest of the proof is identical to the one for $\Log = \NFL$. Let $\Log = \NQFLm$. For the same reasons as for $\NQFL$, for each $b \in \mathsf{PAR}(\B)$, $b = \imath x \varphi \notin \B$. Then, by the definition of $\sim$, ${\D_\E}_\B^\Log$, and ${\I_\B^\Log}_{v_\B^\Log}$, ${\I_\B^\Log}_{v_\B^\Log}(\imath x \varphi)\notin {\D_\E}_\B^\Log$. We conduct the rest of the proof similarly to the one for $\Log \in \{\NFL,\NQFL\}$.\medskip

	\noindent $\psi:=P(t_1,\ldots,t_n)$\quad Let $t_1,\ldots,t_n \in \mathsf{TERM}(\B)$ and $P(t_1,\ldots,t_n) \in \B$. Let $\Log \in \{\PFL,\PQFL\}$. By the definition of $\I_B^\Log$ for predicates and ${\I_B^\Log}_{v_\B^\Log}$,\linebreak $\langle {\I_B^\Log}_{v_\B^\Log}(t_1), \ldots, {\I_B^\Log}_{v_\B^\Log}(t_n)\rangle \in \I_\B^\Log(P)$. By the satisfaction condition for atomic predicate formulas we obtain $\M_\B^\Log, v_\B^\Log \models P(t_1,\ldots,t_n)$ as required. Now let $\Log \in \{\NFL,\NQFL\}$. By expandedness of $\B$ we know that the rule $(\E I_1)$ (\NFL) or $(\E I_1)$ together with $(\E I_3)$ (\NQFL) was applied to $P(t_1,\ldots,t_n)$, thus yielding $\E t_1,\ldots, \E t_n \in \B$. By the proof of the case $\psi := \E t$ we know that\linebreak ${\I_\B^\Log}_{v_\B^\Log} (t_1), \ldots, {\I_\B^\Log}_{v_\B^\Log} (t_n) \in {\D_\E}_\B^\Log$. Moreover, by the definition of $\I_\B^\Log$ and ${\I_\B^\Log}_{v_\B^\Log}$, $\langle {\I_\B^\Log}_{v_\B^\Log}(t_1),\ldots,{\I_\B^\Log}_{v_\B^\Log} (t_n)\rangle \in \I_\B^\Log(P)$. Hence, by the satisfaction condition for atomic predicate formulas, $\M_\B^\Log, v_\B^\Log \models P(t_1,\ldots,t_n)$. Finally, let $\Log = \NQFLm$. By expandedness of $\B$ the rule $(= I_1)$ was applied to $P(t_1, \ldots, t_n)$, thus yielding $a_i=t_i$, for $1 \leq i \leq 2$ and $t_i$ being a definite description. Without loss of generality assume that for each $t_i$, $1\leq i\leq n$, $t_1$ is a definite description, and so, a fresh parameter $a_i$ occurred on $\B$, for $1\leq i\leq n$. By the definition of $\sim$ and ${\D_\E}_\B^\Log$ for $\Log = \NQFLm$ we get that $[t_1]_\sim=[a_1]_\sim,\ldots,[t_n]_\sim=[a_n]_\sim \in {\D_\E}_\B^\Log$. By the definition of $\I_\B^\Log$ and ${\I_\B^\Log}_{v_\B^\Log}$, ${\I_\B^\Log}_{v_\B^\Log}(t_1),\ldots, {\I_\B^\Log}_{v_\B^\Log}(t_n) \in {\D_\E}_\B^\Log$ and $\langle {\I_\B^\Log}_{v_\B^\Log} (t_1),\ldots, {\I_\B^\Log}_{v_\B^\Log} (t_n)\rangle \in \I_\B^\Log(P)$. Hence, by the satisfaction condition for atomic predicate formulas, $\M_\B^\Log, v_\B^\Log \models P(t_1,\ldots,t_n)$.\medskip
	
	\noindent $\psi:=\neg P(t_1,\ldots,t_n)$\quad Let $t_1,\ldots,t_n \in \mathsf{TERM}(\B)$ and $\neg P(t_1,\ldots,t_n) \in \B$. By the definition of $\I_B^\Log$ for predicates and ${\I_B^\Log}_{v_\B^\Log}$, $\langle {\I_B^\Log}_{v_\B^\Log}(t_1), \ldots, {\I_B^\Log}_{v_\B^\Log}(t_n)\rangle \notin \I_\B^\Log(P)$, for otherwise it would mean that $P(t_1,\ldots,t_n) \in \B$, which would contradict the assumption about openness of $\B$. By the satisfaction condition for atomic predicate formulas we obtain $\M_\B^\Log, v_\B^\Log \not\models P(t_1,\ldots,t_n)$ and by the satisfaction condition for $\neg$-formulas we get $\M_\B^\Log, v_\B^\Log \models \neg P(t_1,\ldots,t_n)$ as required.\medskip

	\noindent $\psi:=\forall x\varphi$\quad Let $\forall x\varphi \in \B$. Let $\Log \in \{\PFL,\NFL\}$. By expandedness of $\B$ the rule $(\forall E_2)$ was applied yielding $\varphi[x/b]$ for each $b$ such that $\E b \in \B$. By the inductive hypothesis, for each $b\in\mathsf{TERM}(\B)$ such that ${\I_\B^\Log}_{v_\B^\Log}(b) \in {\D_\E}_\B^\Log$, $\M_\B^\Log , v_\B^\Log \models \varphi[x/b]$. By the satisfaction condition for $\forall$-formulas and \href{lem::Substitution}{Substitution Lemma} we get $\M_\B^\Log, v_\B^\Log \models \forall x\varphi$. Let $\Log \in \{\PQFL, \NQFL\}$. By expandedness of $\B$ the rule $(\forall E_1)$ was applied yielding $\varphi[x/b]$ for each $b\in\mathsf{TERM}(\B)$. To each $b \in \mathsf{TERM}(\B)$ the rule $(\E I_3)$ was applied, so by the inductive hypothesis, for each $b\in\mathsf{TERM}(\B)$ such that ${\I_\B^\Log}_{v_\B^\Log}(b) \in {\D_\E}_\B^\Log$, $\M_\B^\Log , v_\B^\Log \models \varphi[x/b]$. By the satisfaction condition for $\forall$-formulas we get $\M_\B^\Log, v_\B^\Log \models \forall x\varphi$. Let $\Log = \NQFLm$. By expandedness of $\B$ the rule $(\forall E_1)$ was applied yielding $\varphi[x/b]$ for each $b\in\mathsf{TERM}(\B)$. By the definition of ${\D_\E}_\B^\Log$ and by the inductive hypothesis, for each $b\in\mathsf{TERM}(\B)$ such that ${\I_\B^\Log}_{v_\B^\Log}(b) \in {\D_\E}_\B^\Log$, $\M_\B^\Log , v_\B^\Log \models \varphi[x/b]$. By the satisfaction condition for $\forall$-formulas we get $\M_\B^\Log, v_\B^\Log \models \forall x\varphi$.\medskip
	
	\noindent $\psi:=\neg\forall x\varphi$\quad Let $\neg\forall x\varphi \in \B$. Let $\Log \in \{\PFL,\NFL\}$. By expandedness of $\B$ the rule $(\neg\forall E_2)$ was applied yielding $\neg\varphi[x/a]$ and $\E a$ for a certain $a \in \mathsf{PAR}$. By the inductive hypothesis, for some $a\in\mathsf{TERM}(\B)$ such that ${\I_\B^\Log}_{v_\B^\Log}(a) \in {\D_\E}_\B^\Log$, $\M_\B^\Log , v_\B^\Log \models \neg\varphi[x/a]$. By the satisfaction condition for $\forall$-formulas we get $\M_\B^\Log, v_\B^\Log \not\models \forall x\varphi$ and finally, by the satisfaction condition for $\neg$-formulas, we obtain $\M_\B^\Log , v_\B^\Log \models \neg \forall x \varphi$. Let $\Log \in \{\PQFL, \NQFL\}$. By expandedness of $\B$ the rule $(\neg\forall E_1)$ was applied yielding $\neg\varphi[x/a]$ for a certain $a\in\mathsf{TERM}(\B)$. The rule $(\E I_3)$ was applied to $a$ yielding $\E a \in \B$, so by the inductive hypothesis, for some $a\in\mathsf{TERM}(\B)$ such that ${\I_\B^\Log}_{v_\B^\Log}(b) \in {\D_\E}_\B^\Log$, $\M_\B^\Log , v_\B^\Log \models \neg\varphi[x/b]$. By the satisfaction condition for $\forall$-formulas we get $\M_\B^\Log, v_\B^\Log \not\models \forall x\varphi$, and so, by the satisfaction condition for $\neg$-formulas, we obtain $\M_\B^\Log , v_\B^\Log \models \neg\forall x\varphi$. Let $\Log = \NQFLm$. By expandedness of $\B$ the rule $(\neg\forall E_1)$ was applied yielding $\neg\varphi[x/a]$ for a certain $a\in\mathsf{TERM}(\B)$. By the definition of ${\D_\E}_\B^\Log$ and by the inductive hypothesis, for some $a\in\mathsf{TERM}(\B)$ such that ${\I_\B^\Log}_{v_\B^\Log}(a) \in {\D_\E}_\B^\Log$, $\M_\B^\Log , v_\B^\Log \models \neg\varphi[x/a]$. By the satisfaction condition for $\forall$-formulas we get $\M_\B^\Log, v_\B^\Log \not\models \forall x\varphi$, so finally, by the satisfaction condition for $\neg$-formulas, we obtain $\M_\B^\Log , v_\B^\Log \models \neg \forall x\varphi$.\medskip
	
	\noindent $\psi:= \chi \land \theta$ \quad Let $\chi \land \theta \in \B$. By expandedness of $\B$ $(\land E)$ was applied returning $\chi,\theta \in \B$. By the inductive hypothesis $\M_B^\Log, v_\B^\Log \models \chi,\theta$. By the satisfaction condition for $\land$-formulas we obtain $\M_\B^\Log , v_\B^\Log \models \chi \land \theta$.\medskip
	
	\noindent $\psi:= \neg(\chi \land \theta)$ \quad Let $\neg(\chi \land \theta) \in \B$. By expandedness of $\B$ $(\neg\land E)$ was applied returning $\neg\chi\in\B$ or $\neg\theta \in \B$. Assume that the former is the case. By the inductive hypothesis $\M_B^\Log, v_\B^\Log \models \neg\chi$. By the satisfaction condition for $\land$-formulas we obtain $\M_\B^\Log , v_\B^\Log \models \neg(\chi \land \theta)$. If we assume that $\neg\theta \in \B$, the proof is conducted analogously.\medskip
	
	\noindent $\psi:= \neg\neg\chi$ \quad Let $\neg\neg\chi \in \B$. By expandedness of $\B$ $(\neg\neg E)$ was applied returning $\chi \in \B$. By the inductive hypothesis $\M_B^\Log, v_\B^\Log \models \chi$. By the satisfaction condition for $\neg$-formulas applied twice we obtain $\M_\B^\Log , v_\B^\Log \models \neg\neg\chi$.
	\end{proof}

\end{subappendices}

\end{document}